\documentclass[%
reprint,
superscriptaddress,
amsmath,amssymb,
aps,
prb,
floatfix,
]{revtex4-1}
\usepackage{epsfig}
\usepackage{epstopdf}
\usepackage{graphicx}
\usepackage{amsfonts,amsmath,latexsym,amssymb}

\begin{document}

\def\be{\begin{equation}}
\def\ee{\end{equation}}
\def\ua{\uparrow}
\def\da{\downarrow}

\title{Entropy as a function of magnetisation for a 2D spin-ice model exhibiting a Kasteleyn transition}

\author{S. A. Grigera}

\affiliation{SUPA, School of Physics and Astronomy, University of St Andrews, North Haugh, St Andrews KY16\ 9SS, UK}
\affiliation{Instituto de F\'{\i}sica de L\'{\i}quidos y Sistemas  Biol\'ogicos, UNLP-CONICET, La Plata 1900, Argentina}

\author{C. A. Hooley}

\affiliation{SUPA, School of Physics and Astronomy, University of St Andrews, North Haugh, St Andrews KY16\ 9SS, UK}

\date{\today}

\begin{abstract}
We present a combined analytical and numerical study of the entropy as a function of magnetization for an orientable 2D spin-ice model that exhibits a Kasteleyn transition.  The model that we use is related to the well known six-vertex model but, as we show, our representation of it is more convenient for constructing approximate expressions for the entropy at fixed magnetization.  We also discuss directions for further work, including the possibility of deforming our model into one exhibiting a quantum Kasteleyn transition.
\end{abstract}
\maketitle

\section{Introduction}
A popular version of the third law of thermodynamics is that the entropy density of a physical system tends to zero in the $T \to 0$ limit\cite{Mandl2013}.  However, there is a class of theoretical models that violate this law\cite{Bernal1933,Pauling1935,Nagle1966,Lieb1967,Wannier1950,Chow1987,Bramwell2001,Castelnovo2008}:\ models in this class exhibit a ground-state degeneracy which grows exponentially with the system size, leading to a non-zero entropy density even at $T=0$.  Nor can these be easily dismissed as theorists' abstractions, since one also sees ample evidence in experiment
\cite{Giauque1936,Harris1997,Ramirez1999,Higashinaka2003} that there are systems in which the entropy density plateaus at a non-zero value over a large range of temperature.  In many such cases it is suspected that it eventually falls to zero at a much lower temperature scale\cite{denHertog2000,Pomaranski2013}, though recent theoretical work on skyrmion magnets suggests that this might not always be the case\cite{Douccot2016}.

Whatever the ultimate low-temperature fate of these materials, it is clear that over a broad range of temperatures they exhibit physics which is well captured by models with a non-zero residual entropy density.  One important class of these are so-called ice models, in which the ground-state manifold consists of all configurations which satisfy a certain local `ice rule' constraint\cite{Baxter2007}.

The first such model was Pauling's model for the residual configurational entropy of water ice\cite{Pauling1935}.  Here the local constraint is that two of the four hydrogens neighboring any given oxygen should be chemically bonded to it to form a water molecule.  Similar models were subsequently conjectured to apply to the orientations of spins along local Ising axes in magnetic pyrochlore lattices \cite{Anderson1956,Harris1997}, which by analogy were lately dubbed `spin ice' compounds.  Such models develop power-law spin-spin correlations at low temperatures but they do not order.  Their low-temperature state is often referred to as a `co-operative paramagnet' \cite{Villain1979}.

One interesting feature of such co-operative paramagnets is their response to an applied magnetic field.  The configurations that make up the ice-rule manifold usually have different magnetizations; thus an applied field, depending on its direction, may either reduce\cite{Udagawa2002,Moessner2003}
or entirely eliminate\cite{Jaubert2008} the degeneracy.  In the latter case, further interesting physics may arise when the system is heated, especially if the ice-rule constraints do not permit the thermal excitation of individual flipped spins.  In such cases the lowest-free-energy excitation may be a {\it string\/} of flipped spins extending from one side of the system to the other.  A demagnetization transition mediated by such excitations is known as a {\it Kasteleyn transition}\cite{Kasteleyn1963,Jaubert2008}.

An important quantity in such systems is the entropy density of the system as a function of its magnetisation density.  We present in this paper a version of a 2D spin ice model \cite{Wills2002,Chern2012,Wan2012} that is related to the six-vertex model, but for which the calculation of the entropy density may be carried out in a physically transparent fashion in terms of a `Kasteleyn line' representation.  We perform such a calculation, and check our formula against the results of Monte Carlo simulations, showing good agreement up to an overall scaling factor, which we discuss.  This analysis paves the way for the extension of our model to include a transverse magnetic field, thereby allowing the exploration of quantum Kasteleyn physics.

The remainder of this paper is structured as follows.  In section \ref{sec:model}, we present the spin ice model that we shall study, discuss its relationship to other models in the literature, and provide some analytical and numerical results on its thermodynamic properties in the absence of an applied magnetic field.  In section \ref{sec:kasteleyn}, we consider the Kasteleyn transition that the model exhibits in the presence of an applied longitudinal field, and we introduce a `string representation' in terms of which the relevant states may be easily described and counted.  In section \ref{sec:entropy}, we employ this string representation to estimate the model's entropy density as a function of its magnetization density analytically, comparing our results to those of Monte Carlo simulations.  Finally, in section \ref{sec:summary}, we summarize our findings and discuss possible future lines of work.

\section{The model}
\label{sec:model}
\begin{figure}
\centerline{\includegraphics[width=0.95\columnwidth]{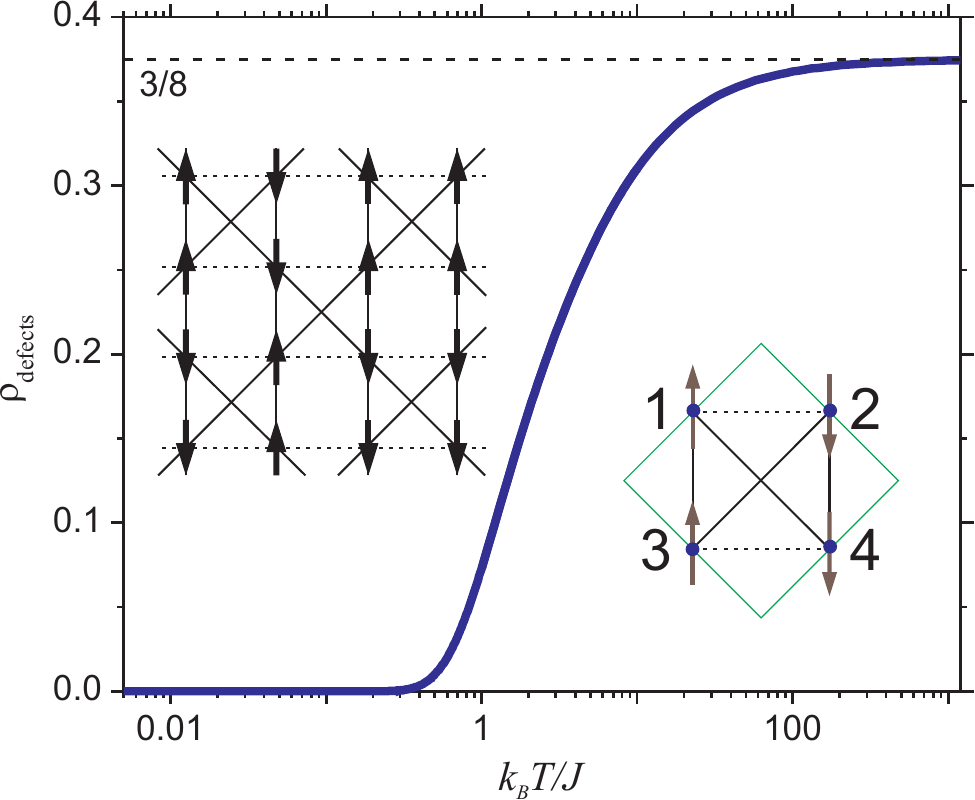}}
\caption{The density of defects, $\rho_{\rm defects}$, as a function of scaled temperature, $k_B T/J$, for a lattice of 8192 spins and in the absence of an applied magnetic field.  The number of defects on a plaquette is defined as the number of single spin-flips by which it differs from the nearest ice-rule configuration.  Thus each state in the ground-state manifold of the system has $\rho_{\rm defects}=0$. The dotted line marks the high-temperature asymptotic value of $3/8$ (see text). Inset (top left): A portion of the lattice, with ferromagnetic bonds represented by solid lines and antiferromagnetic bonds by dotted lines.  Inset (bottom right): The unit cell of the lattice, including the numbering convention we use for the spins on a single plaquette.}
\label{defects}
\end{figure}

The model that we shall consider has the following Hamiltonian:
\be
H = \sum_{ij} J_{ij} \sigma_i \sigma_j - h \sum_i \sigma_i.
\label{ham}
\ee
Here $i$ and $j$ label the sites of a two-dimensional square lattice, $\sigma_i = \pm 1$ is an Ising variable on lattice site $i$, and $h$ is an externally applied (longitudinal) magnetic field.  The exchange interaction $J_{ij}$ is given by:
\be
J_{ij} = \left\{ \begin{array}{lll}
\phantom{-}J & \qquad & {\bf r}_j = {\bf r}_i + {\hat {\bf x}}; \\
-J & \qquad & {\bf r}_j = {\bf r}_i + {\hat {\bf y}}; \\
-J & \qquad & {\bf r}_i = n {\hat {\bf x}} + m {\hat {\bf y}} \,\,\,\,(n+m\,\,\mbox{odd}) \\
& & \quad \mbox{and}\,{\bf r}_j = {\bf r}_i + {\hat {\bf x}} + {\hat {\bf y}}; \\
-J & \qquad & {\bf r}_i = n {\hat {\bf x}} + m {\hat {\bf y}} \,\,\,\,(n+m\,\,\mbox{even}) \\
& & \quad \mbox{and}\,{\bf r}_j = {\bf r}_i - {\hat {\bf x}} + {\hat {\bf y}}; \\
\phantom{-}0 & & \mbox{otherwise,}
\end{array}
\right.
\label{exchanges}
\ee
where ${\bf r}_i$ is the position vector of site $i$, ${\hat {\bf x}}$ and ${\hat {\bf y}}$ are the unit vectors of a Cartesian system in the two-dimensional plane, and $J$ is a positive constant.  In words, this says that the interaction between the spins is antiferromagnetic if the sites are nearest neighbors in the horizontal direction, ferromagnetic if they are nearest neighbors in the vertical direction, ferromagnetic for certain next-nearest-neighbor pairs (those linked by the diagonal solid lines in the upper left inset of Fig.~\ref{defects}), and zero otherwise.

In this paper, we shall always work in the limit $J \gg \vert h \vert, k_B T$.  Furthermore, where necessary we shall take the number of sites in the lattice to be $N$, always assuming $N$ to be large enough that edge effects can be neglected.  When we refer to the density of something (e.g.\ the entropy density), we shall always mean that quantity divided by the number of spins --- not, for example, by the number of plaquettes.

The lattice described by (\ref{exchanges}) is shown in the upper-left inset of Fig.~\ref{defects}, with ferromagnetic bonds represented by solid lines and antiferromagnetic bonds represented by dotted lines.  One may view this lattice as made of corner-sharing plaquettes, one of which is shown in the lower-right inset of Fig.~\ref{defects}.  It is easy to see that the bonds on this plaquette cannot all be satisfied at once:\ the model (\ref{ham}) is therefore magnetically frustrated.

\begin{figure}
\centerline{\includegraphics[width=0.85\columnwidth]{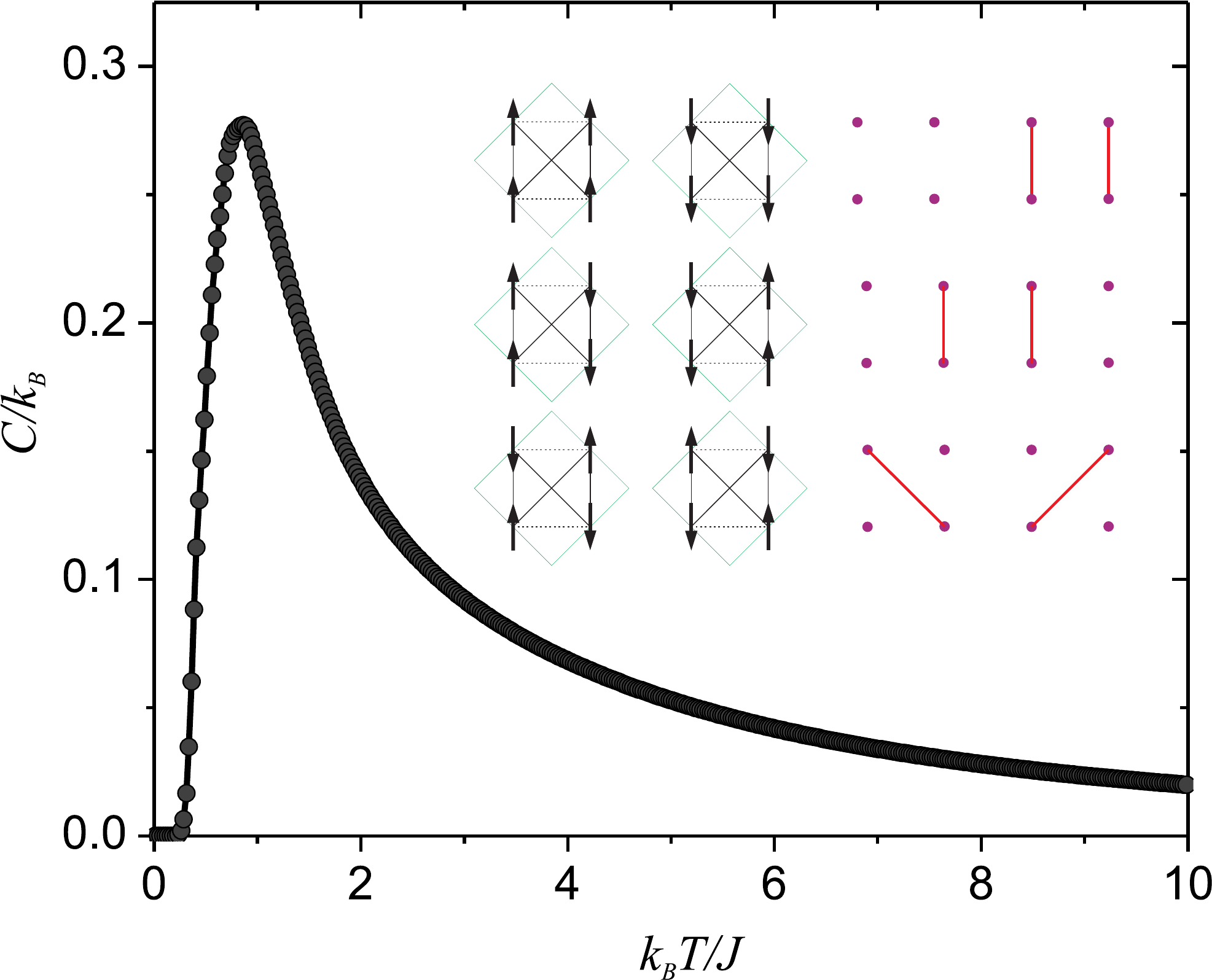}}
\caption{The dimensionless heat capacity per spin, $C/k_B$, as a function of scaled temperature, $k_B T/J$, in the absence of an applied magnetic field, calculated using the Wang-Landau method.  Inset (left):\ the six degenerate zero-field ground states for a single plaquette.  Inset (right):\ the same states in the string representation.}
\label{C}
\end{figure}

The sixteen spin configurations of the elementary plaquette, together with their energies, are shown in Table \ref{plaqconf}.
\begin{table}
\begin{center}
\begin{tabular}{c|c|c|c|c|c|c}
Configuration & $\ua\ua\ua\ua$ & $\ua\da\da\ua$ & $\ua\da\ua\da$ & $\da\ua\da\ua$ & $\da\ua\ua\da$ & $\da\da\da\da$ \\ \hline
Energy & $-2J-4h$ & $-2J$ & $-2J$ & $-2J$ & $-2J$ & $-2J+4h$
\end{tabular}

\vspace*{3mm}
\begin{tabular}{c|c|c|c|c|c|c|c|c}
Configuration & $\ua\ua\ua\da$ & $\ua\ua\da\ua$ & $\ua\da\ua\ua$ & $\da\ua\ua\ua$ & $\da\da\da\ua$ & $\da\da\ua\da$ & $\da\ua\da\da$ & $\ua\da\da\da$ \\ \hline
Energy & $-2h$ & $-2h$ & $-2h$ & $-2h$ & $2h$ & $2h$ & $2h$ & $2h$
\end{tabular}

\vspace*{3mm}
\begin{tabular}{c|c|c}
Configuration & $\ua\ua\da\da$ & $\da\da\ua\ua$ \\ \hline
Energy & $6J$ & $6J$
\end{tabular}
\end{center}
\caption{The energies of the sixteen spin configurations of the elementary plaquette.  Each configuration is specified by listing the orientations of the four plaquette spins in the order corresponding to the numbering in Fig.~\ref{defects}.  The first six configurations listed are those that, in the absence of an external magnetic field, constitute the sixfold-degenerate ground-state (or `ice rule') manifold.}
\label{plaqconf}
\end{table}
When $h=0$, i.e.\ in the absence of an external magnetic field, there are six degenerate ground-state configurations.  They are shown in the left-hand inset of Fig.~\ref{C}:\ we shall call them the `ice-rule configurations,' and the manifold spanned by them the `ice-rule manifold.'

This model is related to others in the literature by various transformations of the spin variables.  For example, if we reverse the sign of each even-numbered row of spins, and simultaneously reverse the sign of the exchange integral on every vertical or diagonal bond, we obtain a model with all antiferromagnetic bonds.  The price we pay is that the magnetic field is now staggered, changing sign from one row of spins to the next:\ thus the resulting model is the antiferromagnetic checkerboard model in a staggered magnetic field.  A further mapping from a global Ising axis to local easy axes\cite{Moessner1998} maps it to the six-vertex model in a vertical electric field\cite{Yang1967}.  However, the advantage of our version of the model lies in the especially simple picture it provides of the exponentially many states in the ice-rule manifold and of the associated Kasteleyn transition.   

Because of these exponentially many ice-rule states, our model does not order as the temperature is reduced.  Rather, it crosses over into a co-operative paramagnetic state in which every plaquette is in one of the ice-rule configurations.  The density of defects (a measure of how many plaquettes are not in an ice-rule configuration) vanishes smoothly as the temperature tends to zero, and the specific heat shows a corresponding Schottky-like peak at temperatures $T \sim J/k_B$ but no sharp features.

Because the ground-state degeneracy is exponential in the system size, the model has a non-zero entropy density even at zero temperature.  A na{\"\i}ve estimate would suggest a value of $k_B \ln 6$ per plaquette, i.e.\ $\frac{1}{2} k_B \ln 6 \approx 0.896\,k_B$ per spin, due to the six-fold ground-state degeneracy.  This estimate, however, is too na{\"\i}ve, since it ignores the important constraint that the ice-rule configurations chosen for two neighboring plaquettes must agree on the orientation of the spin at their shared corner.

We may easily improve our estimate of the zero-temperature entropy density by taking this constraint into account at a local level.  Imagine `growing' a spin configuration of the lattice from top to bottom.  Each time a new row is added, the orientations of spins 1 and 2 of each plaquette of the row being added ($j$) will be fixed by the (already chosen) configuration of the row above ($j-1$).  The ice rules for this model do not favor any particular spin direction for any single site on the plaquette; hence the probabilities of the four configurations of this pair of spins are simply $P_{\ua\ua} = P_{\ua\da} = P_{\da\ua} = P_{\da\da} = 1/4$.  The number of ice-rule configurations consistent with these constraints is (see Fig.~\ref{C}) $N_{\ua\ua} = N_{\da\da} = 1$; $N_{\ua\da} = N_{\da\ua} = 2$.  Thus half the plaquettes in the new row have no choice of configuration, while the other half may choose between two.  This gives an average entropy per plaquette of $\frac{1}{2} k_B \ln 2$, which corresponds to an entropy density of $\frac{1}{4} k_B \ln 2 \approx 0.173\,k_B$ per spin.

This estimate is still rather crude, since it neglects correlations between the configurations of neighboring plaquettes in row $j-1$, which will be induced by their connections to a common plaquette in row $j-2$.  However, it was shown by Lieb \cite{Lieb1967} that such correlation corrections may be resummed to yield an exact result for the ground-state entropy density of such `square ice' models:\ $s_0 \equiv S_0/N = \frac{3}{4} k_B \ln \left( \frac{4}{3} \right) \approx 0.216\,k_B$.  We shall call this value the `Lieb entropy density,' and denote it $s_0^{\rm Lieb}$.

\begin{figure}
\centerline{\includegraphics[width=\columnwidth]{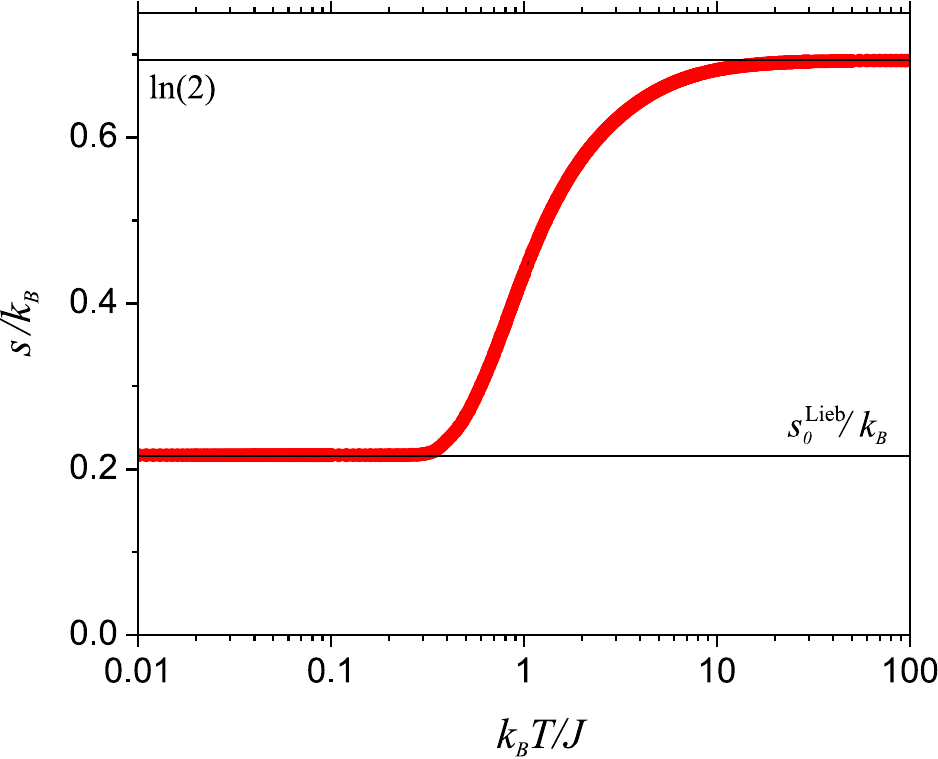}}
\caption{The dimensionless entropy density of the system, $s/k_B$, as a function of scaled temperature, $k_B T/J$, for a lattice of 8192 spins and in the absence of an applied magnetic field, calculated using the Wang-Landau method.  At high temperatures the entropy density is that of an Ising paramagnet, $k_B \ln 2$ per spin.  The zero-temperature residual entropy density is consistent with Lieb's exact result for two-dimensional ice models, $s_0^{\rm Lieb} = \frac{3}{4} k_B \ln \left( \frac{4}{3} \right) \approx 0.216\,k_B$.}
\label{Entropy}
\end{figure}

All of the above expectations are borne out by Monte Carlo simulations of the model, the results of which are shown in Figs.~\ref{defects}--\ref{C}.

First, we demonstrate the increasing predominance of ice-rule configurations as the temperature is lowered.  For this it is useful to define the number of defects on a plaquette as the number of single spin-flips by which the spin configuration deviates from the closest ice-rule configuration.  By this measure, the states in the top line of Table \ref{plaqconf} have zero defects, those in the second line have one, and those in the third line have two.  Fig.~\ref{defects} shows the density of defects as a function of temperature.
%

The asymptotic high-temperature value of this quantity can be easily calculated.  In the infinite-temperature limit all configurations of a plaquette are equally probable, i.e.\ each has a probability $\frac{1}{16}$.  From Table \ref{plaqconf}, we see that there are six configurations with no defects, eight configurations with one, and two configurations with two.  Hence the average number of defects per plaquette at infinite temperature is $0 \times \frac{6}{16} + 1 \times \frac{8}{16} + 2 \times \frac{2}{16} = \frac{3}{4}$.  Since there are twice as many spins as plaquettes, the defect density is simply half of this, i.e.\ $\rho_{\rm defects} \to \frac{3}{8} = 0.375$ as $k_B T/J \to \infty$.

Second, we calculate the entropy density of the system as a function of temperature, using the Wang-Landau method\cite{WangLandau2001a}.  The results are shown in Fig.~\ref{Entropy}.  At high temperatures the entropy density tends to $k_B \ln 2$, the Ising paramagnetic value.  At low temperatures it tends to a non-zero constant value which is in good agreement with the Lieb entropy density $s_0^{\rm Lieb}$ given above.  In between there are no sharp features, confirming that the model exhibits only a crossover from high-temperature paramagnetic to low-temperature cooperative-paramagnetic behavior.

Third, we obtain the specific heat capacity as a function of temperature, also using the Wang-Landau method.  The results are shown in Fig.~\ref{C}.  In keeping with our results for the entropy density in Fig.~\ref{Entropy}, we see that although there is a broad Schottky-like peak at temperatures of order $J/k_B$ there are no sharp features, supporting our expectation that this model would not exhibit a phase transition.
\section{String representation}
\label{sec:kasteleyn}
The particular magnetization distribution of the states in our model's ice-rule manifold gives it an unusual response to an externally applied longitudinal magnetic field.  In this section, we discuss the Kasteleyn transition that results from this, and use it to motivate a `string representation' of the ice-rule states that will be useful in calculating an approximate expression for the entropy as a function of magnetization.

We call the longitudinal magnetic field $h$, and in the following we shall take it to be positive.  As shown in the first line of Table \ref{plaqconf}, the degeneracy between the six ice-rule configurations is lifted as soon as the field $h$ is applied.  Indeed, for any non-zero $h$ (and remembering that we always work in the $h \ll J$ limit) the ground state of a plaquette is the unique `all up' configuration.  It follows that, at $T=0$, the entire lattice simply has $\sigma_i = +1$ for all sites $i$.

Now let us consider what happens to this fully magnetized state as the temperature is increased.  One might expect the appearance of a dilute set of `down' spins.  However, a feature of this model is that a single spin-flip takes the system out of the ice-rule manifold, and at $h,k_B T \ll J$ this will not occur.  To understand what will happen instead, let us introduce a representation of the states in the ice-rule manifold in terms of strings.

We begin with a single plaquette.  If we take as our reference state the one in which all the spins are up, we may represent the six ice-rule configurations in terms of lines joining the spins that are down.  This is shown in the right-hand inset of Fig.~\ref{C}.  Representing the `all down' configuration as two vertical lines rather than two horizontal ones is in principle arbitrary, but it has the advantage of yielding a model in which these lines of down spins can neither cross each other nor form closed loops.

\begin{figure}
\vspace{-5mm}
\centerline{\includegraphics[width=1.1\columnwidth]{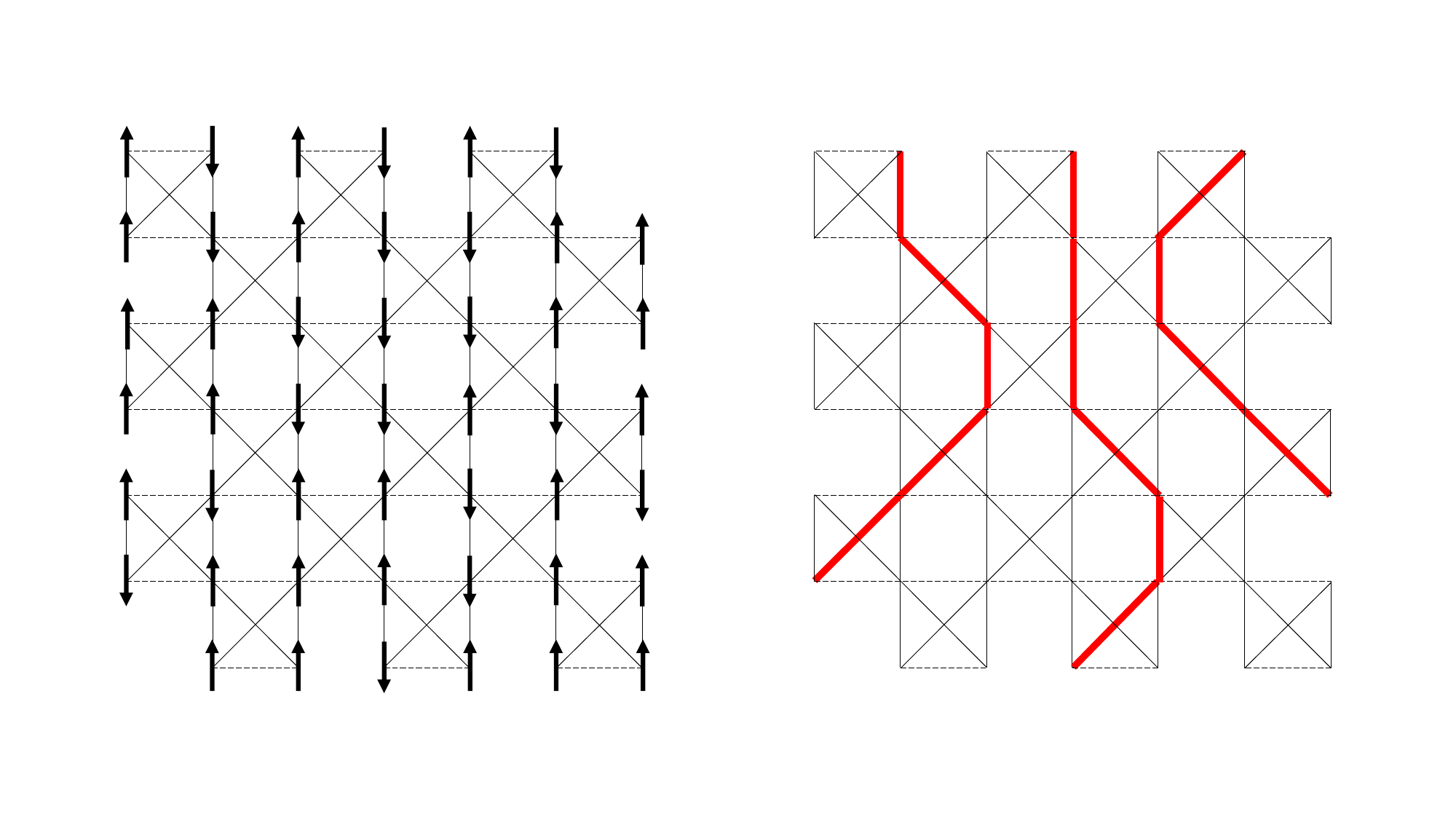}}
\vspace{-8mm}
(a) \hspace{3.84cm} (b)
\caption{An example of the mapping between ice-rule configurations of the spins and the string picture.  (a) A particular ice-rule spin configuration.  (b) The same configuration in the string representation, obtained by mapping each plaquette as shown in the right-hand inset of Fig.~\ref{C}.}
\label{spinstring}
\end{figure}

To make an ice-rule-obeying configuration of the entire lattice, we must put these plaquettes together in such a way that any string that leaves one plaquette enters its neighbor.  Thus there is a one-to-one mapping between ice-rule-obeying configurations of the spins $\sigma_i$ and configurations of these strings.  Each string must extend all the way across the lattice.  An example of such a mapping is shown in Fig.~\ref{spinstring}, where panel (a) shows a portion of the lattice in a particular ice-rule spin configuration, and panel (b) shows the same configuration represented in terms of strings of `down' spins.  Notice that the strings cannot loop back on themselves:\ there is no plaquette in the right-hand inset of Fig.~\ref{C} for which the string is horizontal, and the two types of diagonal line cannot join into a `V' shape because plaquettes on the same row of the lattice do not share corners.  Notice also that the strings cannot cross:\ a plaquette containing all `down' spins is always to be interpreted as a pair of vertical strings, not as a pair of diagonally crossing ones, in order to preserve the one-to-one nature of the mapping.

To proceed further, let us suppose that the lattice consists of $L_x$ sites in the horizontal direction and $L_y$ sites in the vertical direction, so that $N = L_x L_y$.  Each string, irrespective of its configuration, contains precisely $L_y$ spins, so that a configuration with $N_s$ strings has $N_s L_y$ down spins and thus an energy of $2 h N_s L_y$ relative to the fully magnetized state (or `string vacuum').  Such a string is the {\it minimal\/} demagnetizing excitation of the system that is consistent with the ice rule.

Since a single string has an energy cost proportional to the linear size of the system, it might appear that such strings cannot be thermally excited.  This is not true, however, because a single string also has two choices about which way to go every time it enters a new plaquette, meaning that its entropy of $k_B L_y \ln 2$ is also proportional to $L_y$.  Thus the free-energy cost of introducing a single string into the fully magnetized state is
\be
F = E - TS = \left( 2h - k_B T \ln 2 \right) L_y.
\ee
When the temperature reaches the critical value $T_c = 2h/(k_B \ln 2)$, this free-energy cost flips sign, and the system becomes unstable to the proliferation of strings.  (This is somewhat similar to what happens in a Berezinskii-Kosterlitz-Thouless transition\cite{Berezinskii1971,Kosterlitz1973}, except that in our model we do not have `positive' and `negative' strings, so the physics of screening plays no r{\^o}le.)

In fact the increase in the string density from zero for $T > T_c$ --- which corresponds directly to the decrease in the magnetization from its saturated value --- is continuous.  This is because the above argument applies strictly only to a single string introduced into the fully magnetized state.  Once a finite density of strings has been created the entropy associated with new ones is reduced, and thus the temperature at which it becomes free-energetically favorable to create them goes up.

This kind of transition, in which the elementary thermal excitations are system-spanning strings, is called a {\it Kasteleyn transition\/}.  It was first described by Kasteleyn in the context of dimer models\cite{Kasteleyn1963}.

\begin{figure}
\centerline{\includegraphics[width=\columnwidth]{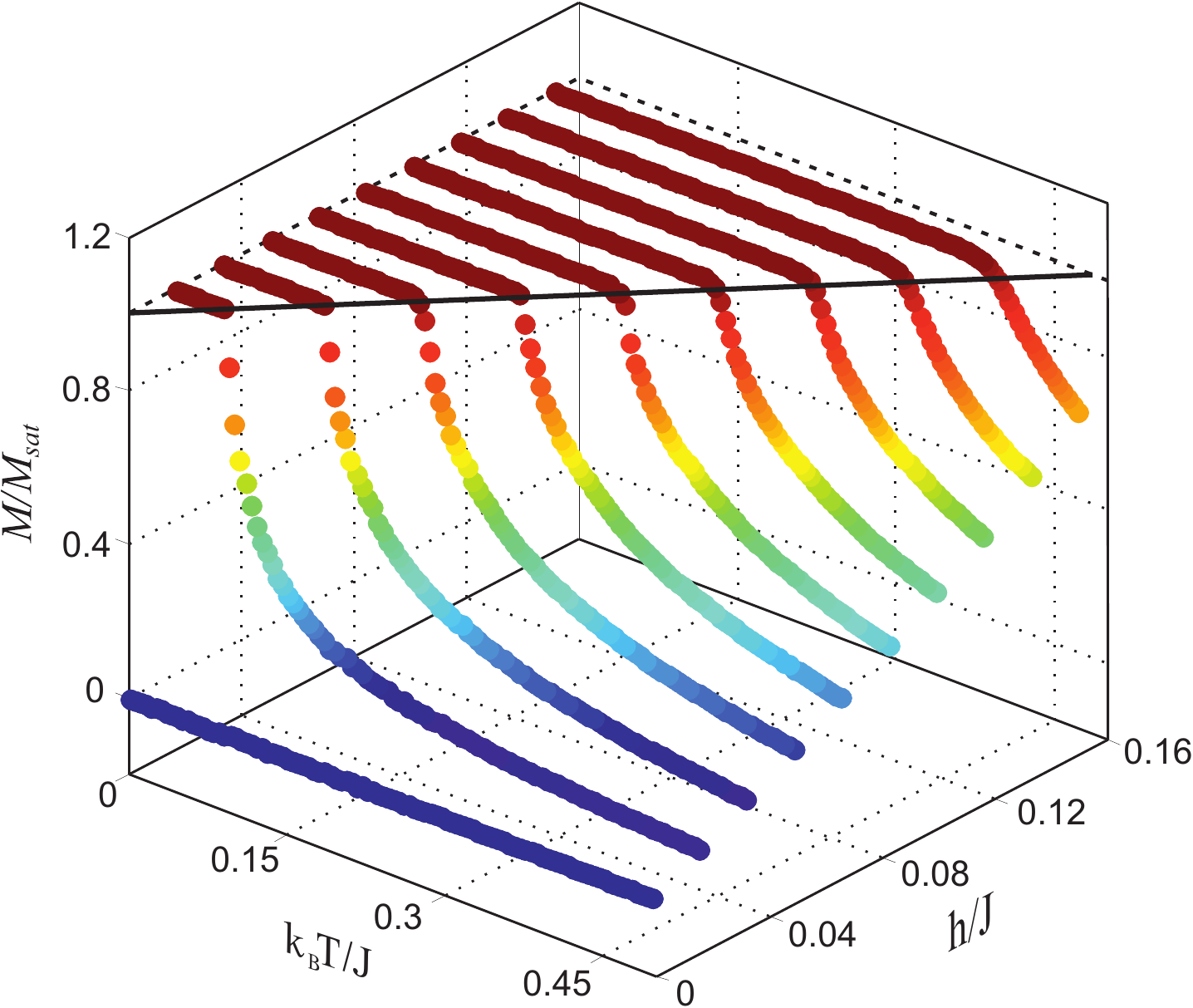}}
\caption{The ratio of the magnetization to its saturated value, $M/M_{\rm sat}$, as a function of scaled temperature, $k_B T/J$, and scaled longitudinal field, $h/J$.  The solid black line shows the theoretical prediction for the Kasteleyn transition temperature, $T_c = 2h/(k_B \ln 2)$.}
\label{MvsHT}
\end{figure}

\begin{figure}
\centerline{\includegraphics[width=\columnwidth]{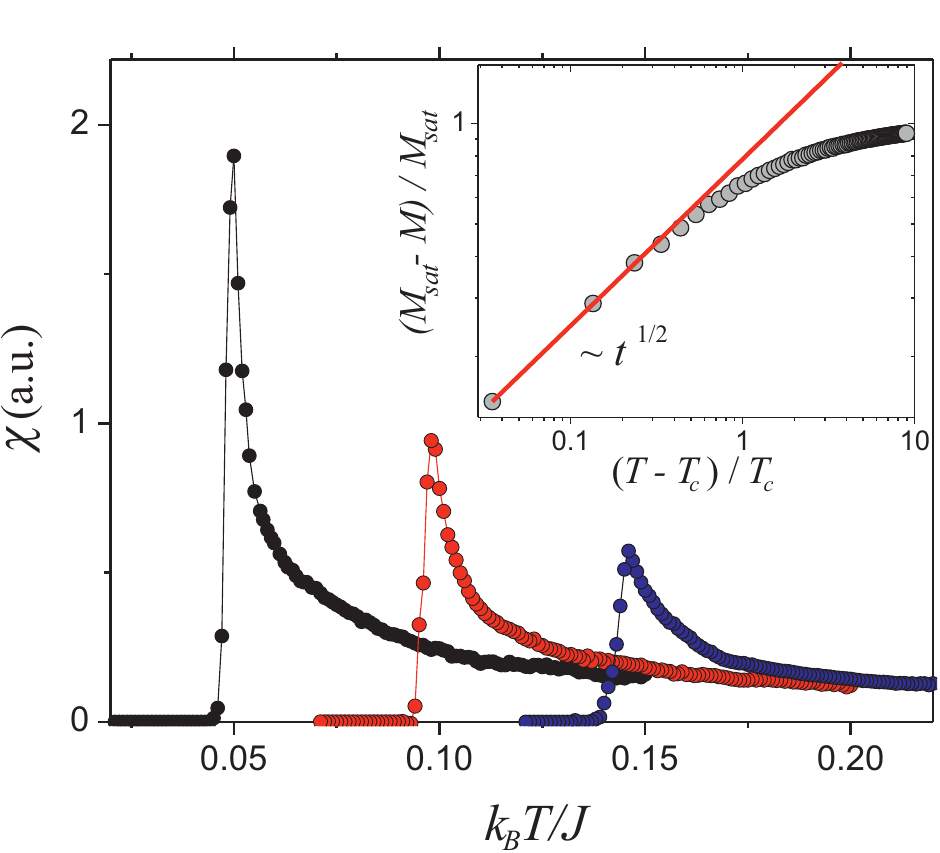}}
\caption{The magnetic susceptibility, $\chi$, as a function of scaled temperature, $k_B T/J$, for a lattice of 8192 spins with three different values of the scaled magnetic field $h/J$: 0.017 (black symbols, leftmost peak), 0.034 (red symbols, middle peak), and 0.051 (blue symbols, rightmost peak). The inset shows the reduced magnetization, $\mu$, as a function of the reduced temperature, $t$, for an applied field $h/J=0.017$ (grey filled circles).  The solid red line corresponds to $\mu \sim t^{1/2}$.}
\label{chivsT}
\end{figure}

\begin{figure}
\centerline{\includegraphics[width=\columnwidth]{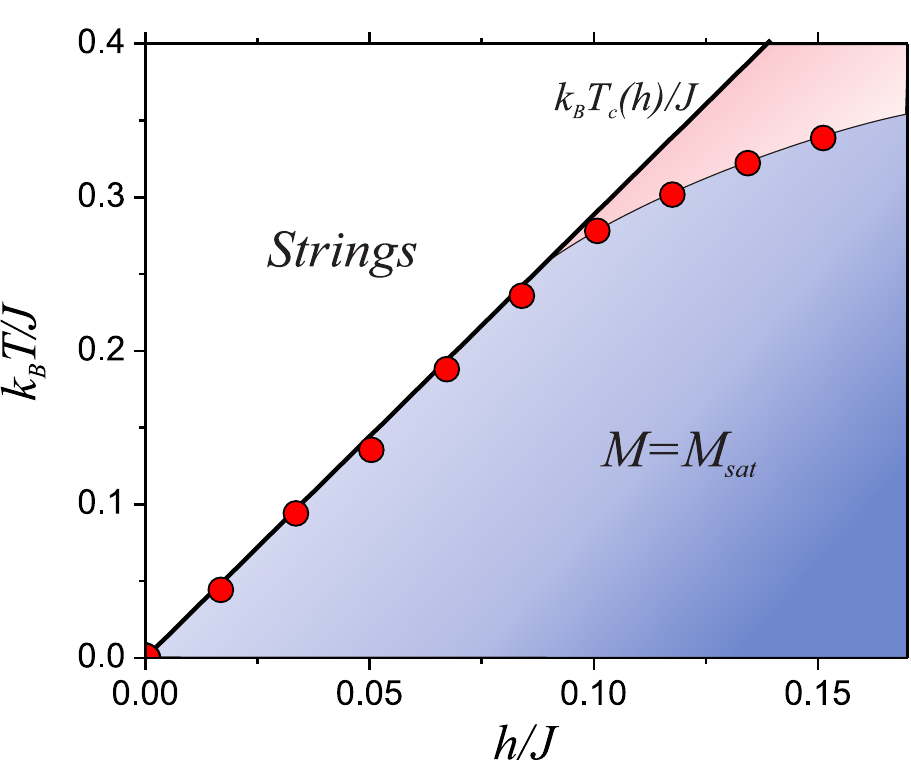}}
\caption{The phase diagram of our model as a function of scaled temperature, $k_B T/J$, and scaled magnetic field, $h/J$.  The red dots show the Kasteleyn temperature as determined from the magnetization curves, i.e.\ the temperature at which the magnetization first departs from its saturated value.  The black line is the theoretical prediction $T_c(h)= 2h/(k_B \ln 2)$.  As expected, the simulation results depart from the theoretical prediction at temperatures where the condition that the spin configuration remain strictly in the ice-rule manifold, $h, k_B T\ll J$, is no longer fulfilled (pink area).}
\label{TkvsH}
\end{figure}

The above predictions are again borne out by our Monte Carlo simulations, the results of which are shown in Figs.~\ref{MvsHT}--\ref{TkvsH}.

Fig.~\ref{MvsHT} shows a three-dimensional plot of the equilibrium value of the magnetization, $M$, as a function of the temperature and the applied magnetic field.  At all temperatures below $T_c(h)$ the magnetization takes its saturated value; above $T_c(h)$ it decreases smoothly with increasing temperature, tending to zero only as $T \to \infty$.  This may be understood in the string representation of the problem.  As more and more strings are introduced, the entropy density of each new one decreases; in the limit where half the lattice sites are populated by strings it tends to zero, meaning that this will occur only in the infinite-temperature limit.

Fig.~\ref{chivsT} shows the magnetic susceptibility, determined at three different values of the applied field.  In each case, one sees at $T=T_c(h)$ the asymmetric peak characteristic of a Kasteleyn transition.  This highlights an intriguing consequence of the physics of the Kasteleyn strings:\ below $T_c(h)$ the linear susceptibility is strictly zero, while as $T_c(h)$ is approached from above the susceptibility diverges.

For a two-dimensional Kasteleyn transition one expects to find $\beta= 1/2$ on the high-temperature side\cite{Nagle1975,Moessner2003}, that is,
\be
\mu \sim t^{1/2},
\ee
where $\mu \equiv (M_{\rm sat}-M)/M_{\rm sat}$ is the reduced magnetization and $t \equiv (T - T_c)/T_c$ is the reduced temperature.  This is indeed the case in our simulations:\ the inset of Fig.~\ref{chivsT} is a log-log plot of $\mu$ as a function of $t$, calculated for a system of 8192 spins and with an applied field of $h/J = 0.017$ (grey filled circles), compared with the expected $t^{1/2}$ behavior (solid red line).  Similar behavior is found for all simulated fields below $0.1 h/J$. 

In Fig.~\ref{TkvsH} we collect our data into a phase diagram.  The filled red circles show the temperature of the Kasteleyn transition, determined from the data in Fig.~\ref{MvsHT} as the temperature at which the magnetization departs from its saturated value.  The thick black line is the prediction $T_c(h) = 2h/(k_B \ln 2)$ derived above.  The departure of the red points from this line at larger fields and temperatures is due to the violation of the condition $h, k_B T \ll J$.  In the pink region the thermal excitations are not full strings, but instead string fragments extending from one ice-rule-violating plaquette to another.  The physics of such string fragments, and their signatures in neutron scattering, were discussed by Wan and Tchernyshyov\cite{Wan2012}.
%
\section{Entropy as a function of magnetization}
\label{sec:entropy}
In this section, we come to the main point of our paper:\ to use the string representation to calculate the entropy density of the system, $s$, at a fixed value of the magnetization density, $m \equiv M/M_{\rm sat}$.  Clearly $s(m)$ is an even function of $m$, so we may restrict our calculation to the case $m \geqslant 0$.
The magnetization density may equivalently be expressed as the density of strings, $\eta_s$, via the formula $\eta_s = (1-m)/2$.

To determine the entropy density corresponding to a given value of $\eta_s$, consider propagating the string configuration downwards from the top of the lattice.  We shall assume that this propagation has reached a certain row $j$, and concentrate on a single string in that row.  As it enters a new plaquette in row $j+1$, this string has in principle two choices:\ to continue vertically downwards, or to cross the plaquette diagonally.  However, if another string is entering the same plaquette, it has only one choice, since the strings cannot cross (see Fig.~\ref{C}).

The probability that a second string enters the same plaquette in row $j+1$ as the first is simply $\eta_s$.  Thus the average number of choices available to the first string upon entering the new plaquette is $\eta_s \times 1 + (1-\eta_s) \times 2 = 2 - \eta_s$.  This means that each string has a total entropy $S_s \approx k_B L_y \ln \left( 2 - \eta_s \right)$; with a total number of strings $\eta_s L_x$, it follows that the total entropy is $S \approx k_B L_x L_y \eta_s \ln \left( 2 - \eta_s \right)$.  Dividing by the number of spins $N=L_x L_y$, and using $\eta_s = (1-m)/2$, we obtain
\be
s_0(m) \approx {\tilde s}_0(m) \equiv k_B \left( \frac{1-m}{2} \right) \ln \left( \frac{3+m}{2} \right). \label{sofm}
\ee

\begin{figure}
\centerline{\includegraphics[width=\columnwidth]{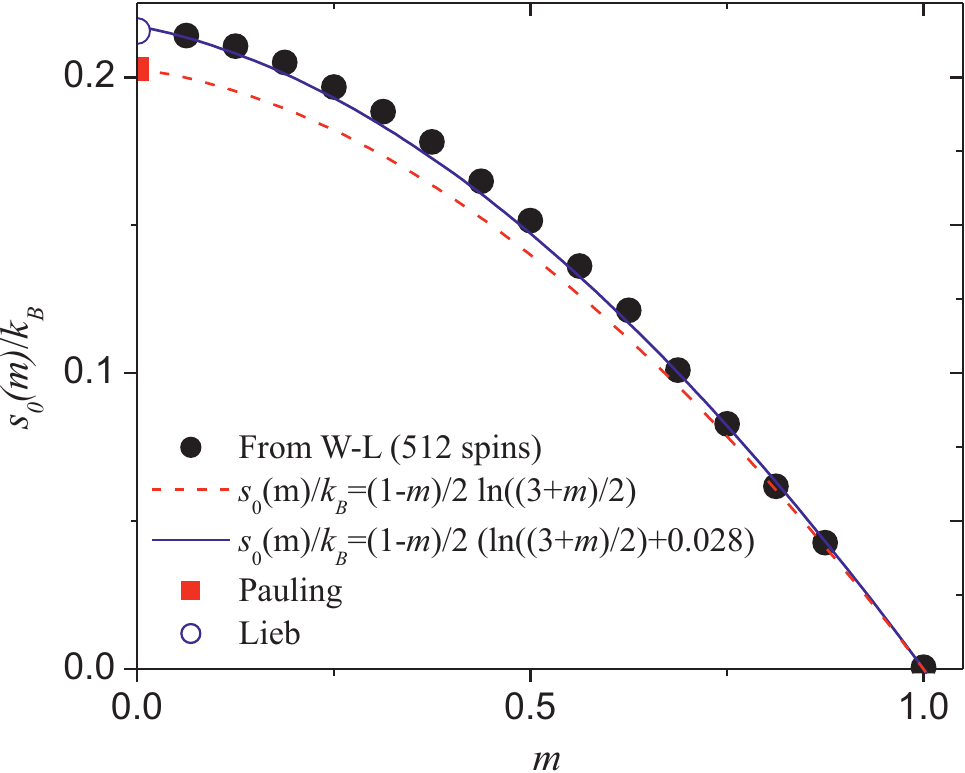}}
\caption{The residual dimensionless entropy per site, $s_0/k_B$, as a function of the scaled magnetization per site, $m \equiv M/M_{\rm sat}$.  The black filled circles show the values obtained numerically using the Wang-Landau method for a lattice of 512 spins.  The dashed red line is the free-string result ${\tilde s}_0(m)$ (see text); note that it tends to Pauling's entropy at $m=0$ (filled red square).  The solid blue line is the curve obtained by multiplying the number of microstates by a constant factor, chosen to rescale ${\tilde s}_0(0)$ to match Lieb's exact result (open blue circle).}
\label{S0vsM}
\end{figure}

In Fig.~\ref{S0vsM} we compare this approximation with numerical results for the entropy density obtained using the Wang-Landau method.  The filled black circles are the numerical results, while the dashed red curve is our analytical approximation (\ref{sofm}).  It is clear that these were never going to coincide, since the $m \to 0$ limit of ${\tilde s}_0(m)$ is the Pauling entropy density, $\frac{1}{2} k_B \ln \left( \frac{3}{2} \right)$, while the $m \to 0$ limit of the actual entropy density is the Lieb entropy density, $\frac{3}{4} k_B \ln \left( \frac{4}{3} \right)$.
%

The origin of the difference between Lieb's exact result and Pauling's approximation lies in positive correlation of closed loops\cite{Nagle1966,Lieb1967}, which increases by a small factor the number of possible configurations obeying the ice rule.  If one makes the crude assumption that this factor is independent of $m$, this results in a constant additive change to the logarithm in (\ref{sofm}):
\be
{\tilde s}_0(m) \longrightarrow k_B \left( \frac{1-m}{2} \right) \left[ \ln \left( \frac{3+m}{2} \right) + \alpha \right].
\ee
If we choose the constant $\alpha$ to match the known result at $m=0$, the resulting curve (shown in blue) gives a very reasonable fit to the numerical data points over the whole range $0 \leqslant m \leqslant 1$.

\vspace{5mm}
\section{Summary and future work}
\label{sec:summary}
In this paper, we have presented a spin-ice model defined on a two-dimensional lattice of mixed ferro- and antiferromagnetic bonds.  We have used its Kasteleyn transition (a known feature of the six-vertex model to which it can be mapped) to motivate the introduction of a `string representation' of the ice-rule manifold, and we have demonstrated that this representation is well adapted to the task of making an analytical estimate of the entropy density as a function of the magnetization density.

One appealing feature of models in this class is that, unlike full three-dimensional spin ices, the Ising quantization axis is the same on each lattice site.  This makes it natural to consider adding to the model a spatially uniform transverse magnetic field, $\Gamma$.  The results of this should be particularly interesting in the $h,\Gamma,k_B T \ll J$ regime, where the applied field is expected to stabilize the string phase at low temperatures, leading to a line of quantum Kasteleyn transitions in the zero-temperature $(h,\Gamma)$ plane.  This extension of the model (\ref{ham}) is the subject of a forthcoming work \cite{dhgb}.

\vspace{1mm}

\begin{acknowledgments}
We are pleased to acknowledge useful discussions with Rodolfo Borzi, Daniel Darroch, and Peter Holdsworth.  This research was supported in part by the National Science Foundation under Grant No.\ NSF PHY11-25915, and CAH is grateful to the Kavli Institute for Theoretical Physics in Santa Barbara for their hospitality.  CAH also gratefully acknowledges financial support from the EPSRC (UK) via grant number EP/I031014/1.  SAG would like to acknowledge financial support from CONICET, and ANPCYT (Argentina) via grant PICT-2013-2004.
\end{acknowledgments}

\appendix*
\section{Numerical methods}

We performed Monte Carlo simulations using the Metropolis\cite{Metropolis1953} and Wang-Landau\cite{WangLandau2001a,WangLandau2001b} algorithms. In
both cases we used a single-spin flip algorithm on systems of $L \times L$ unit cells (see inset of Fig.\ \ref{defects}) ranging from $L=8$ (128 spins) to $L=64$  (8192 spins). For our Metropolis algorithm, we used $5 \times 10^3$
Monte Carlo steps for equilibration and $L^{-2} \times 10^8$ for averaging.  In order to calculate the entropy of the system we used the Wang-Landau algorithm to determine the density of states, $\delta$. We labeled the states according to their energy, $E_i$, and magnetization, $M_i$. For normalization we used the condition $\sum_{E_i,M_i}\delta(E_i,M_i) = 2^N$, where $N$ is the total number of spins of the system.  The modification factor changed from $\ln(f_0)=1$ to $\ln(f_{\rm final})=10^{-9}$.


%

\end{document}